\documentclass[letterpaper]{article} 
\usepackage[preprint]{aaai2027} 
\usepackage[hyphens]{url} 
\usepackage{graphicx} 
\usepackage{natbib} 
\usepackage{caption} 
\usepackage{booktabs}
\usepackage{amsmath}
\usepackage{amssymb}

\title{ReliGRec: Reliability-Oriented LLM-Based Generative Recommendation via User-Risk-Aware Prompt Routing}
\author{
Haoran Yang\textsuperscript{\rm 1},
Fei Chen\textsuperscript{\rm 1},
Yutian Xiao\textsuperscript{\rm 2},
Jiahao Liang\textsuperscript{\rm 3}
}
\affiliations{
\textsuperscript{\rm 1}Central South University, Changsha, China\\
\textsuperscript{\rm 2}Beihang University, Beijing, China\\
\textsuperscript{\rm 3}South China University of Technology, Guangzhou, China
}

\begin{document}
\maketitle

\begin{abstract}
User behavior in real-world recommender systems is heterogeneous. While some
users exhibit coherent preferences, others show abrupt interest shifts, bursty
interactions, excessive repetition, or inconsistency with collaborative
neighborhoods. Such deviations may arise from benign variation or manipulation,
including shilling attacks, but do not alone establish malicious intent.
Existing robust recommenders exploit user-risk signals through training-time
reweighting or graph aggregation, whereas adapting generation to estimated user
level weak risk remains underexplored in LLM-based generative recommendation.
We propose ReliGRec (Reliability-oriented Generative Recommendation), a weakly
supervised framework whose name denotes its design goal rather than a
supervised reliability variable. ReliGRec derives user-level weak-risk proxy
labels from review-feedback signals for a subset of users and represents
sequential behavior and collaborative context using a Behavior Token and
temporal Graph Tokens, respectively. A Dual-View Weak-Risk Estimator fuses the
representations to produce a user-level weak-risk score that selects a Simple
or Cautious Prompt at inference. The Cautious Prompt is designed to encourage
attention to stable, collaboratively supported evidence while reducing
overreliance on isolated, short-term, or repeated interactions. The Behavior Token affects
generation through weak-risk estimation and routing, whereas the aggregated
Graph Token provides collaborative context for next-item Semantic ID
generation. ReliGRec thus turns weak-risk estimation from an auxiliary
prediction into a generation-time control signal. Experiments report
competitive recommendation and weak-risk proxy-label prediction, while routing
analyses characterize the recommendation-quality and inference-cost behavior
of weak-risk-guided prompting.

\end{abstract}

\section{Introduction}
\begin{figure}[t]
\centering
\includegraphics[width=\columnwidth]{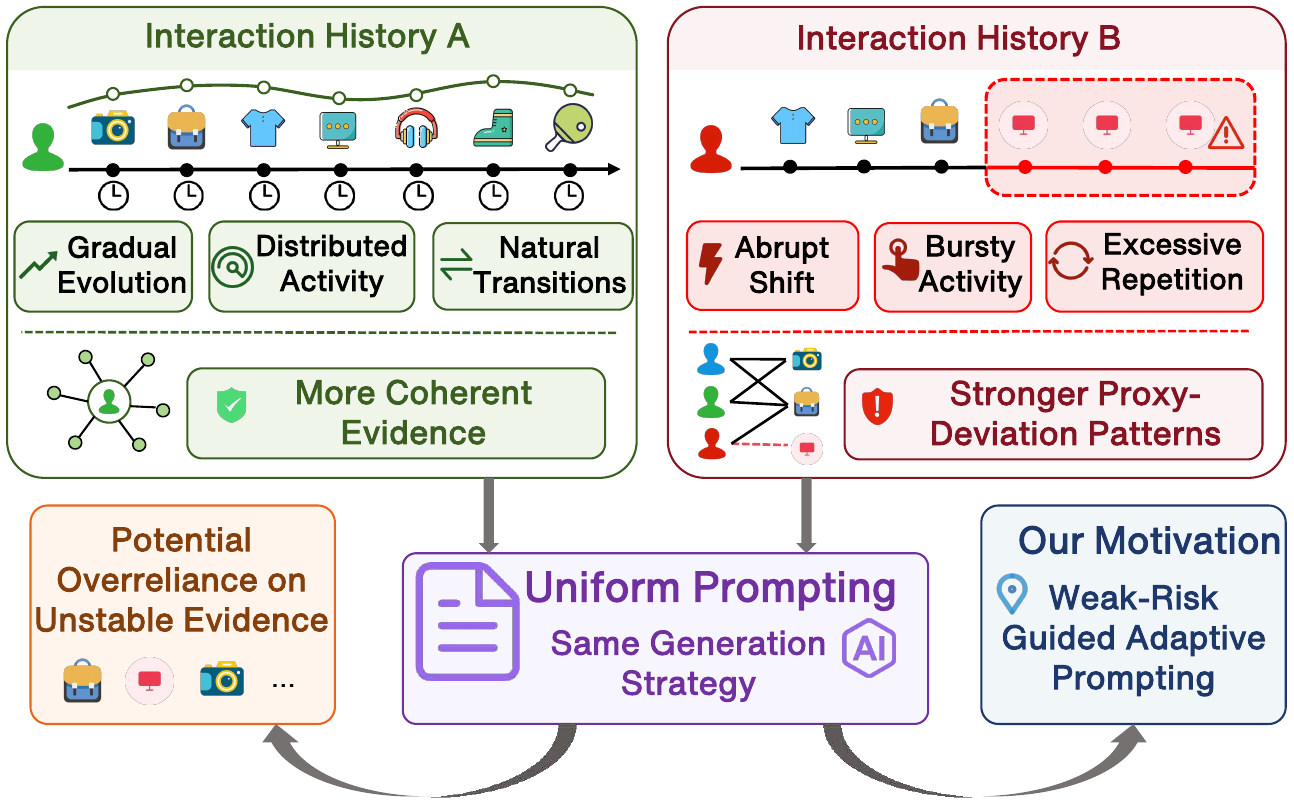}
\caption{Motivation for user-level weak-risk-guided prompt routing. Uniform
prompting applies the same generation policy to users with different behavioral
and collaborative deviation patterns, potentially overemphasizing unstable
evidence.}
\label{fig:intro-motivation}
\end{figure}

Traditional recommenders typically score and rank candidate items through
sequential modeling or user--item graph
learning~\citep{kang2018sasrec,he2020lightgcn,fan2021tgsrec}. More recently,
generative recommendation has emerged as a complementary paradigm that
predicts the next item by autoregressively generating item
identifiers~\citep{rajput2023tiger}. With large language models (LLMs)
demonstrating strong semantic understanding and autoregressive generation
capabilities, recent studies have further explored LLMs as generative
backbones, representing items as discrete Semantic ID sequences and
incorporating collaborative, multimodal, or pretrained semantic
knowledge~\citep{zheng2024lcrec,wang2024letter,zhang2025collm,
hu2026gencdr,ye2026align3gr,zhao2026musicrec,mu2026graphlora}. However, such
LLM-based generation remains strongly conditioned on user interaction
histories. Real histories may contain abrupt preference shifts, bursty
activity, excessive repetition, feedback noise, or collaborative inconsistency
arising from benign variation or manipulation such as shilling
attacks~\citep{zhang2020graphrfi,zhang2024lorec,nguyen2025pgt4rec}. A uniform
prompt may therefore cause the LLM to over-rely on unstable historical
evidence. Figure~\ref{fig:intro-motivation} motivates estimating user-level
weak risk from observed interaction histories before decoding and using it to
adapt the LLM's generation strategy.

Existing work improves recommendation reliability at three stages.
Training-side robust recommenders use risk signals for reweighting, graph
aggregation, or representation rectification, as in GraphRfi, LoRec, PGT4Rec,
Trust-GRS, and
DITaR~\citep{zhang2020graphrfi,zhang2024lorec,nguyen2025pgt4rec,
mu2025trustgrs,qin2026ditar}, but their interventions mainly affect learning
rather than inference-time generation. Generation-side methods improve item
tokenization, behavior modeling, collaborative injection, or inductive
inference through TIGER, LC-Rec, LETTER, CoLLM, MusicRec, Align$^3$GR, GenCDR,
GraphLoRA, and
SpecGR~\citep{rajput2023tiger,zheng2024lcrec,wang2024letter,zhang2025collm,
zhao2026musicrec,ye2026align3gr,hu2026gencdr,mu2026graphlora,ding2026specgr},
without explicitly adapting decoding to user-level weak risk inferred from
observed interaction histories. Output-side approaches such as GUIDER and UGR
use generation uncertainty for reranking, diversity adjustment, rejection, or
truncation~\citep{xu2026guider,fan2026ugr}, but do not select a pre-decoding
generation policy based on user-level weak risk. Thus, connecting user-level
weak-risk estimation with generation control while retaining dynamic
collaborative evidence remains underexplored.

This objective poses two challenges. First, proxy deviation cannot be reliably
identified from individual behavioral cues. A short-term preference shift or
repeated interaction may reflect either normal preference evolution or a higher
proxy-deviation pattern. The key challenge is therefore to capture user-level
deviation patterns from observed interactions and estimate their relative
proxy-deviation tendency under partial weak supervision. Behavioral evidence
alone may be insufficient for this distinction and should be complemented by
temporal collaborative context. We define the resulting signal as user-level
weak risk, which is estimated from a user's observed interaction history and
weakly supervised by user-level weak-risk proxy labels derived from
review-feedback signals. This signal characterizes deviation from regular
interaction patterns rather than verified malicious status or a calibrated
attack probability. Second, the estimated weak risk should inform generation
rather than remain an isolated auxiliary prediction, while avoiding unnecessary
intervention for users with lower proxy deviation. This calls for a lightweight
mechanism that adjusts how historical evidence is used before decoding.

To address these challenges, we propose ReliGRec, a weakly supervised
risk-aware generative recommendation framework. ReliGRec uses a Behavior Token
and temporal Graph Tokens to encode ordered interactions and evolving
collaborative structure, respectively. Their fusion in a Dual-View Weak-Risk
Estimator mitigates the ambiguity of single-view evidence. The resulting
user-level weak-risk score selects either a Simple Prompt or a Cautious Prompt
for each user. The Simple Prompt directly uses the available history and
collaborative context, whereas the Cautious Prompt is designed to discourage
overreliance on isolated, transient, or excessively repetitive evidence and to
prioritize temporally stable and collaboratively supported patterns. The
aggregated Graph Token also conditions next-item Semantic ID generation,
enabling dynamic collaborative evidence to support both user-level weak-risk
estimation and recommendation generation.

Our main contributions are threefold:
\begin{itemize}
  \item[$\bullet$] We investigate user-level weak-risk-aware generation control
  based on observed interaction histories, using weak risk to select a
  prompting policy before decoding and distinguishing it from training-time
  risk weighting and output-side uncertainty handling.

  \item[$\bullet$] We develop ReliGRec by integrating behavioral and temporal
  collaborative representations, dual-view user-level weak-risk estimation,
  Simple/Cautious Prompt routing, and Graph Token-conditioned Semantic ID
  generation.

  \item[$\bullet$] We evaluate ReliGRec on Beauty and Yelp, reporting
  recommendation and weak-risk proxy-label prediction and characterizing the
  quality--efficiency behavior of weak-risk-guided prompt routing.
\end{itemize}

\section{Problem Definition}
Let $\mathcal U$ and $\mathcal I$ denote the user and item sets, respectively.
For user $u$, the observed interaction prefix up to position $t$ is

\begin{equation}
 \mathcal H_u^{(t)}=(i_{u,1},\ldots,i_{u,t}),
\label{eq:history-prefix}
\end{equation}

with $i_{u,t+1}$ as the next-item prediction target. Each item $i$ is
represented by a length-$L$ Semantic ID:

\begin{equation}
 \operatorname{SID}(i)=(c_{i,1},\ldots,c_{i,L}),
\label{eq:item-sid}
\end{equation}

following the item-identifier formulation of generative
retrieval~\citep{rajput2023tiger}.

We study user-level weak-risk estimation with partially labeled users, using
weak supervision derived from review-feedback signals. Let
$\mathcal U_L\subset\mathcal U$ denote the
labeled user set. For $u\in\mathcal U_L$, $y_u\in\{0,1\}$, where $y_u=0$ and
$y_u=1$ denote the lower- and higher-proxy-deviation groups, respectively; the
remaining users are unlabeled. These labels characterize deviations from
regular interaction patterns rather than verified malicious identities.

\section{Methodology}
\begin{figure*}[t]
  \centering
  \includegraphics[width=\textwidth]{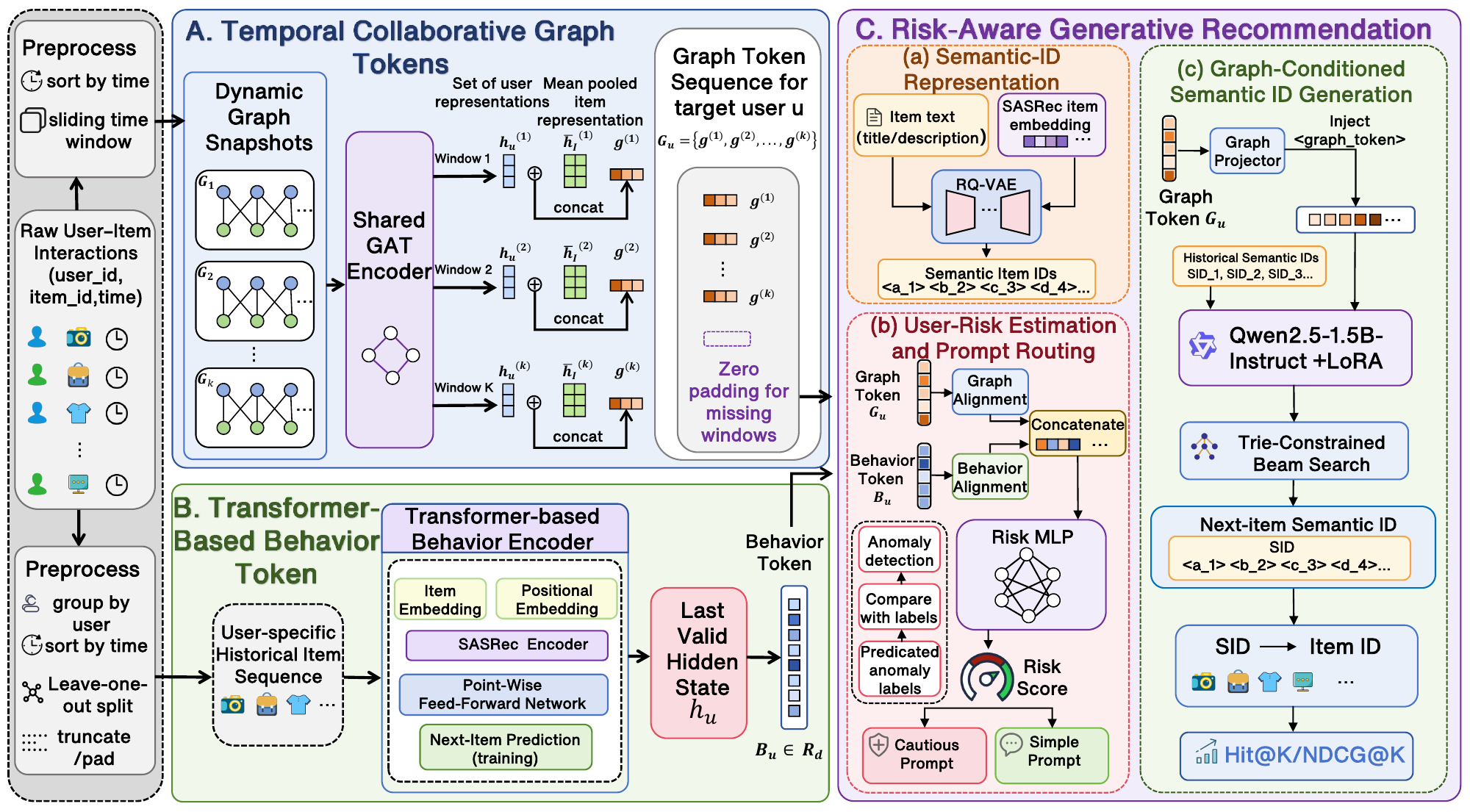}
  \caption{Overall framework of ReliGRec. The Behavior Token and temporal
  Graph Tokens jointly support user-level weak-risk estimation, whose output
  controls prompt routing. The aggregated Graph Token further provides dynamic
  collaborative information for next-item Semantic ID generation.}
  \label{fig:framework}
\end{figure*}

As shown in Figure~\ref{fig:framework}, ReliGRec integrates hierarchical
Semantic IDs, a Behavior Token, and temporal Graph Tokens. The two user
representations estimate weak risk for Simple/Cautious Prompt routing. The
selected prompt and projected Graph Token then condition next-item generation,
while the Behavior Token influences generation through risk estimation and
routing.

\subsection{Semantic ID Construction}

Raw item IDs serve only as indices and do not reflect semantic or collaborative
relationships among items. Following the Semantic ID paradigm in generative
recommendation~\citep{rajput2023tiger,wang2024letter}, we employ
RQ-VAE~\citep{lee2022rqvae} as an offline item tokenizer. Given a semantic
feature $\mathbf{x}_i$ extracted from the title and description of item $i$ and
a collaborative embedding $\mathbf{e}_i^{\mathrm{cf}}$ learned from
user--item interactions, the tokenizer is optimized by

\begin{equation}
\mathcal{L}_{\mathrm{sid}}
=
\mathcal{L}_{\mathrm{recon}}
+\lambda_q\mathcal{L}_{\mathrm{quant}}
+\lambda_{\mathrm{cf}}\mathcal{L}_{\mathrm{cf}},
\label{eq:sid-loss}
\end{equation}

where $\mathcal{L}_{\mathrm{recon}}$ and
$\mathcal{L}_{\mathrm{quant}}$ are the reconstruction and residual-quantization
losses, respectively, and $\mathcal{L}_{\mathrm{cf}}$ aligns the quantized
representation with the collaborative embedding; $\lambda_q$ and
$\lambda_{\mathrm{cf}}$ weight the two corresponding terms. RQ-VAE
successively quantizes residuals, and the resulting codebook indices constitute
the hierarchical Semantic ID defined in Equation~\ref{eq:item-sid}. Valid
Semantic IDs and their associated items are stored in a lookup table and used
to represent both historical items and next-item generation targets.

\subsection{Dual-Path User Representation}

ReliGRec constructs temporal Graph Tokens and a Behavior Token in parallel to
represent the user's dynamic collaborative environment and ordered behavior,
respectively.

\subsubsection{Temporal Graph Token Encoding}

Interaction sequences preserve behavioral order but do not explicitly model
evolving collaborative relations. ReliGRec therefore divides the training
interactions into $K$ overlapping windows, constructs a user--item bipartite
graph for each, and employs a shared graph encoder to obtain user states
$\mathbf{h}_u^{(k)}$ and item states $\mathbf{h}_i^{(k)}$
~\citep{fan2021tgsrec,zhang2023graphreview,tao2026tardgr}. The Graph Token for
window $k$ is defined as

\begin{equation}
\begin{aligned}
\mathbf{m}^{(k)}
&=\frac{1}{|\mathcal{I}^{(k)}|}
\sum_{i\in\mathcal{I}^{(k)}}\mathbf{h}_i^{(k)},\\
\mathbf{g}_u^{(k)}
&=[\mathbf{h}_u^{(k)};\mathbf{m}^{(k)}].
\end{aligned}
\label{eq:window-graph}
\end{equation}

Here, $\mathcal{I}^{(k)}$ is the item set in window $k$, and
$\mathbf{m}^{(k)}$ is its aggregated item context. Thus,
$\mathbf{g}_u^{(k)}$ retains both the user's structural state and window-level
collaborative information.

The window representations are then chronologically organized as
$\mathbf{G}_u=(\mathbf{g}_u^{(1)},\ldots,\mathbf{g}_u^{(K)})$. Missing
windows are zero-padded. A temporal Transformer aggregates the fixed-length
sequence, and its output at the final window position is used as the user
representation:

\begin{equation}
\bar{\mathbf{g}}_u
=
\operatorname{LN}\!\left(
\left[
\operatorname{Tr}_{\mathrm{temp}}
(\mathbf{G}_u+\mathbf{P})
\right]_{K}
\right).
\label{eq:temporal-graph}
\end{equation}

Here, $\mathbf{P}$ denotes temporal position embeddings. The resulting
$\bar{\mathbf{g}}_u$ summarizes the temporal evolution of collaborative
structures, providing dynamic evidence for weak-risk estimation and next-item
generation.

\subsubsection{Behavior Token Encoding}

Temporal Graph Tokens capture evolving collaborative relations but may obscure
the fine-grained order of a user's own interactions. ReliGRec therefore
retains the most recent $L_u$ items from the user's training history:

\begin{equation}
\mathcal{S}_u=(i_{u,1},\ldots,i_{u,L_u}).
\label{eq:behavior-sequence}
\end{equation}

After adding item and positional embeddings, the input sequence is represented
as

\begin{equation}
\mathbf{E}_u
=[\mathbf{e}_{i_{u,1}}+\mathbf{p}_1,\ldots,
\mathbf{e}_{i_{u,L_u}}+\mathbf{p}_{L_u}].
\label{eq:behavior-input}
\end{equation}

A Transformer with a padding mask then encodes the sequence, with the last
valid hidden state taken as the Behavior Token:

\begin{equation}
\mathbf{H}_u
=\operatorname{Tr}_{\mathrm{beh}}
(\mathbf{E}_u;\mathbf{M}_u^{\mathrm{pad}}),
\qquad
\mathbf{B}_u=\mathbf{H}_{u,\mathrm{last}}.
\label{eq:behavior-token}
\end{equation}

Here, $\mathbf{e}_{i_{u,j}}$ and $\mathbf{p}_j$ denote item and positional
embeddings, while $\mathbf{M}_u^{\mathrm{pad}}$ excludes padded positions from
attention. The resulting $\mathbf{B}_u$ summarizes the recent interaction
sequence, providing behavioral evidence complementary to the collaborative
representation for weak-risk estimation.

\subsection{Weak-Risk Estimation and Weak-Risk-Guided Prompt Routing}

ReliGRec fuses behavioral and collaborative representations into a user-level
weak-risk score and converts the score into a prompt-routing policy before
decoding.

\subsubsection{Dual-View Weak-Risk Estimator}

Behavioral and temporal collaborative representations may provide different
signals associated with proxy deviation. ReliGRec therefore combines the
Behavior Token and aggregated temporal Graph Token by projecting them into a
shared $d_p$-dimensional space:

\begin{equation}
\mathbf{g}'_u=\operatorname{LN}_g(\mathbf{W}_g\bar{\mathbf{g}}_u),
\qquad
\mathbf{B}'_u=\operatorname{LN}_b(\mathbf{W}_b\mathbf{B}_u).
\label{eq:risk-projection}
\end{equation}

Here, $\mathbf{W}_g$ and $\mathbf{W}_b$ are projection matrices, and
$\operatorname{LN}$ denotes layer normalization. The projected features are
concatenated and passed through a two-layer MLP to obtain the user-level weak-risk
score:

\begin{equation}
a_u=\operatorname{MLP}_{\mathrm{risk}}
([\mathbf{B}'_u;\mathbf{g}'_u]),
\qquad
s_u=\sigma(a_u).
\label{eq:risk-score}
\end{equation}

Here, $s_u\in(0,1)$ measures the agreement between the user features and the
proxy deviation pattern, rather than a calibrated attack probability or a
confirmed malicious judgment. Let $\mathcal{D}_{\mathrm{risk}}\subseteq
\mathcal{D}_{\mathrm{rec}}$ contain the recommendation training instances whose
users have weak-risk labels. Consistent with the implementation, each such
instance inherits its user's label, and the estimator is optimized using
positive-class-weighted binary cross-entropy:

\begin{equation}
\mathcal{L}_{\mathrm{risk}}
\;=\;
\frac{1}{|\mathcal{D}_{\mathrm{risk}}|}
\sum_{(u,t)\in\mathcal{D}_{\mathrm{risk}}}
\ell_{\mathrm{WBCE}}(a_u,y_u),
\label{eq:risk-loss}
\end{equation}
Here, the weighted binary cross-entropy is defined as
\[
\ell_{\mathrm{WBCE}}(a,y)
=-\omega_{+}y\log\sigma(a)
-(1-y)\log\!\left(1-\sigma(a)\right).
\]

where $\omega_{+}=N_{-}/N_{+}$ is computed from the labeled training
instances. Unlabeled users are excluded from the weak-risk loss but remain
available for recommendation training. During inference, $s_u$ determines the
prompt policy.

\subsubsection{Weak-Risk-Guided Prompt Routing}

During training, the weak-risk proxy labels determine prompt assignment: users in the
higher-proxy-deviation group receive the Cautious Prompt, whereas all remaining
users receive the Simple Prompt. At inference, routing is based on the
predicted weak-risk score $s_u$ and a fixed threshold $\tau$:

\begin{equation}
\begin{aligned}
\mathcal{P}_u^{\mathrm{train}}
&=
\begin{cases}
\mathcal{P}_{\mathrm{cau}},&u\in\mathcal{U}_L,\ y_u=1,\\
\mathcal{P}_{\mathrm{simple}},&\text{otherwise},
\end{cases}\\
\mathcal{P}_u^{\mathrm{test}}
&=
\begin{cases}
\mathcal{P}_{\mathrm{cau}},&s_u\ge\tau,\\
\mathcal{P}_{\mathrm{simple}},&s_u<\tau.
\end{cases}
\end{aligned}
\label{eq:prompt-routing}
\end{equation}

Here, $\tau=0.5$ is a fixed operational cutoff rather than a calibrated
posterior threshold. Prompt assignment is discrete: training instances use
their weak-label-assigned prompts, whereas inference uses the predicted score.
Consequently, the recommendation loss does not optimize the weak-risk
estimator through the hard routing decision.

The Simple template asks the model to recommend directly from the history and
Graph Token. The Cautious template marks the input as potentially unreliable
and instructs the model to identify stable and repeated patterns, use the
Graph Token as supplementary context, exclude candidates supported mainly by
short-term or noisy signals, and select one item aligned with longer-term
interests. These instructions specify an intended evidence-use policy rather
than an empirically verified reasoning process or a judgment of malicious
identity. Both prompts impose the same output constraint: exactly one valid
Semantic ID is generated without intermediate analysis.

\subsection{Graph-Conditioned Semantic ID Generation}

To inject dynamic collaborative information into the generator, a two-layer
projector maps the aggregated Graph Token into the LLM hidden space:

\begin{equation}
\mathbf{z}_u^g
=
\mathbf{W}_2\operatorname{GELU}
(\mathbf{W}_1\bar{\mathbf{g}}_u+\mathbf{b}_1)
+\mathbf{b}_2.
\label{eq:graph-projector}
\end{equation}

The projected representation $\mathbf{z}_u^g$ replaces a reserved Graph Token
placeholder in the input embedding sequence. The selected prompt, the Semantic
IDs of the historical items, and $\mathbf{z}_u^g$ are jointly supplied to an
instruction-tuned LLM. The generator is parameter-efficiently fine-tuned with
LoRA and autoregressively predicts the discrete codes at each level, producing
the complete Semantic ID of the next item.

\subsection{Training Objectives and Constrained Inference}

The Semantic ID tokenizer is first trained offline using the objective in
Equation~\ref{eq:sid-loss}. For the main model, given the target Semantic ID
sequence

\begin{equation}
\mathbf{c}_{u,t+1}
=\operatorname{SID}(i_{u,t+1})
=(c_{i_{u,t+1},1},\ldots,c_{i_{u,t+1},L}),
\label{eq:target-sid}
\end{equation}

we append the end-of-sequence token $c_{\mathrm{eos}}$ during teacher forcing
and define the generation context as

\begin{equation}
\mathbf{x}_{u,t}
=
\left(
\operatorname{SID}(\mathcal{H}_u^{(t)}),
\mathbf{z}_u^g,
\mathcal{P}_u^{\mathrm{train}}
\right).
\label{eq:generation-context}
\end{equation}

Here, $\operatorname{SID}(\mathcal{H}_u^{(t)})$ denotes the ordered Semantic ID
sequences of the observed items, and $\mathbf{z}_u^g$ is the projected Graph
Token actually inserted into the LLM input. The recommendation loss is the
mean cross-entropy over all unmasked target tokens:

\begin{equation}
\begin{aligned}
\mathcal{L}_{\mathrm{rec}}
={}&-\frac{1}{N_{\mathrm{rec}}}
\sum_{(u,t)\in\mathcal{D}_{\mathrm{rec}}}
\sum_{\ell=1}^{L+1}
\log p_{\theta}\!\bigl(
\tilde c_{u,t+1,\ell}
\mid{}\\
&\mathbf{x}_{u,t},
\tilde{\mathbf{c}}_{u,t+1,<\ell}
\bigr).
\end{aligned}
\label{eq:rec-loss}
\end{equation}

where $\mathcal{D}_{\mathrm{rec}}$ denotes the recommendation training
instances, $\tilde{\mathbf{c}}_{u,t+1}=(\mathbf{c}_{u,t+1},
c_{\mathrm{eos}})$ is the supervised response, and $N_{\mathrm{rec}}$ is the
number of its unmasked target tokens in the batch. The generation model and
the weak-risk estimator are jointly optimized using

\begin{equation}
\mathcal{L}
=
\mathcal{L}_{\mathrm{rec}}
+\lambda_{\mathrm{risk}}\mathcal{L}_{\mathrm{risk}},
\label{eq:joint-loss}
\end{equation}

where $\lambda_{\mathrm{risk}}$ controls the contribution of weak-risk
estimation to the overall training objective.

At inference, the model first computes the user-level weak-risk score and
selects a prompt. A prefix Trie is then constructed from the valid Semantic ID
sequences of the candidate items~\citep{decao2022mgenre}. At each decoding
step, only tokens that form valid Semantic ID prefixes are retained, and
completed sequences are ranked by the sums of their token log-probabilities.
The generated Semantic IDs are mapped back to concrete items, after which
previously interacted items are filtered to obtain the final recommendation
list.

\section{Experiments}
We evaluate ReliGRec on Beauty and Yelp. Overall recommendation and weak-risk
estimation are reported on both datasets, while controlled component and
routing analyses are conducted on Beauty. The experiments answer the following
research questions:

\begin{itemize}
    \item[$\bullet$] \textbf{RQ1:} How does ReliGRec perform compared with representative
    graph-based, sequential, generative, and risk-aware recommendation methods
    for next-item prediction?

    \item[$\bullet$] \textbf{RQ2:} How accurately does ReliGRec estimate user-level
    weak-risk proxy labels, and how do behavioral and temporal collaborative
    signals contribute to the estimation?

    \item[$\bullet$] \textbf{RQ3:} How does weak-risk-guided prompt routing affect
    recommendation quality and inference efficiency?

    \item[$\bullet$] \textbf{RQ4:} To what extent does aggregated Graph Token conditioning
    contribute to next-item generation?

\end{itemize}

\subsection{Experimental Settings}

\paragraph{Datasets.}
We use Amazon Beauty~\citep{mcauley2015image} and Yelp2018 released with
LightGCN~\citep{he2020lightgcn}. Beauty is processed with 5-core filtering,
while for Yelp we remove interactions without valid timestamps and apply
10-core filtering. Interactions are chronologically ordered and evaluated under
leave-one-out splitting. Dataset statistics are reported in
Table~\ref{tab:data_statistics}.

\begin{table}[t]
\centering
{
\small
\setlength{\tabcolsep}{3.0pt}
\begin{tabular}{lrrrrr}
\toprule
Dataset & \#Users & \#Items & \#Interactions & Avg. Length & Sparsity \\
\midrule
Beauty & 22,363 & 12,101 & 198,502   & 8.88  & 99.93\% \\
Yelp   & 30,545 & 34,139 & 1,387,627 & 45.43 & 99.87\% \\
\bottomrule
\end{tabular}
}
\caption{Statistics of the processed datasets.}
\label{tab:data_statistics}
\end{table}

For weak-risk estimation, review-level voting ratios are defined as
helpful/total for Beauty and useful/(useful+funny+cool) for Yelp. Reviews
without votes are unlabeled; ratios above 0.85 for Beauty (at least 0.85 for
Yelp) indicate lower deviation, while ratios below 0.40 indicate higher
deviation. Users with at least three labeled reviews are assigned by majority
evidence; otherwise, they remain unlabeled. Labeled users are split 8:2 into
user-disjoint training and test sets. Unlabeled users participate only in
recommendation training. Table~\ref{tab:weak_label_statistics} reports the
resulting distribution.

\begin{table}[t]
\centering
{
\small
\setlength{\tabcolsep}{4.0pt}
\begin{tabular}{lrrr}
\toprule
Dataset & Lower & Higher & Unlabeled \\
\midrule
Beauty & 6,980 (31.21\%) & 1,852 (8.28\%) & 13,531 (60.51\%) \\
Yelp   & 9,968 (32.63\%) & 2,390 (7.82\%) & 18,187 (59.54\%) \\
\bottomrule
\end{tabular}
}
\caption{Distribution of user-level weak-risk proxy labels. Lower and Higher
denote the lower- and higher-proxy-deviation groups, respectively. Percentages
are computed over all users in each dataset.}
\label{tab:weak_label_statistics}
\end{table}

\paragraph{Baselines.}
We compare ReliGRec with six representative baselines.
LightGCN~\citep{he2020lightgcn} captures collaborative relations through
simplified graph propagation, while SASRec~\citep{kang2018sasrec} models
sequential dependencies using self-attention. LETTER-LC-Rec and LETTER-TIGER
combine the LETTER tokenizer~\citep{wang2024letter} with
LC-Rec~\citep{zheng2024lcrec} and TIGER~\citep{rajput2023tiger}, respectively,
representing Semantic-ID-based generative recommendation.
GraphRfi~\citep{zhang2020graphrfi} and PGT4Rec~\citep{nguyen2025pgt4rec}
jointly perform recommendation and user anomaly detection. To ensure task
consistency, we adapt their original recommendation objectives to next-item
prediction while retaining their anomaly-detection components. Their
recommendation outputs are included in RQ1, and their protocol-matched
continuous scores for the weak-risk proxy task are included in RQ2. All
baselines are tuned on the validation set under the same experimental protocol.

\paragraph{Evaluation Metrics.}
For next-item recommendation, we report Hit@1, Hit@5, Hit@10, NDCG@5, and
NDCG@10.
For user-level weak-risk estimation, we report AUPRC, AUROC, and F1, with
AUPRC treated as the primary metric because of label imbalance. F1 is
calculated using the fixed routing threshold (\(\tau=0.5\)). We further
evaluate routing efficiency using the average number of input tokens, average
inference latency per test instance, and the proportion of users routed to the
Cautious Prompt. All methods use the same data split, test users, candidate
catalog, history-aware filtering rule, and evaluation procedure.

\paragraph{Implementation Details.}
ReliGRec uses Qwen2.5-1.5B-Instruct~\citep{yang2024qwen25} with
LoRA~\citep{hu2022lora} (rank 16, scaling factor 32, and dropout 0.05). The
two-layer Behavior Encoder has a hidden size of 32 and encodes up to 50
interactions. Temporal graphs use window/stride sizes of 20,000/10,000 for
Beauty and 50,000/25,000 for Yelp. Cached 128-dimensional Graph Tokens are
obtained using a fixed two-head GAT for Beauty and mean aggregation for Yelp,
and are further aggregated by a two-layer temporal Transformer. The weak-risk
estimator is trained with positive-class-weighted binary cross-entropy, with
\(\lambda_{\mathrm{risk}}=1\) and routing threshold \(\tau=0.5\). The item
tokenizer is a four-level RQ-VAE with 256 entries per codebook and
32-dimensional codes. The generator is trained using AdamW, cosine decay,
BF16, and a per-device batch size of 8. The maximum training budgets are five
epochs for Beauty and six epochs for Yelp; checkpoints are selected on the
validation set. Learning rates are \(7\times10^{-4}\) for Beauty and
\(3\times10^{-4}\) for Yelp. During inference, the 20 most recent interactions
are retained, and catalog-constrained beam search with a beam size of 50 is
combined with seen-item filtering. All experiments are conducted on two NVIDIA
RTX PRO 5000 Blackwell GPUs.
All reported model results are point estimates obtained with seed 42; we do
not use the small cross-method differences to claim statistical significance.

\subsection{Overall Performance (RQ1)}

Table~\ref{tab:overall_results} compares the next-item recommendation
performance of ReliGRec and the baselines under the unified evaluation
protocol.

As shown in Table~\ref{tab:overall_results}, ReliGRec achieves the best results
across all five metrics on Beauty. Compared with the strongest baseline,
LETTER-TIGER, it improves H@1, H@5, H@10, N@5, and N@10 by 54.67\%, 19.24\%,
8.93\%, 27.28\%, and 19.29\%, respectively. On Yelp, ReliGRec obtains an H@1 of 0.00540,
tying with LETTER-TIGER for the best result, and ranks second on the other four
metrics. Its gaps from the corresponding best results range from only 0.12\%
to 1.87\%. Overall, ReliGRec shows a clear advantage on Beauty while remaining
comparable to the strongest baseline on Yelp under the reported protocol.
Because Table~\ref{tab:overall_results} compares complete systems, these
results do not isolate the effects of routing, graph conditioning, or
tokenizer and backbone differences.

\begin{table*}[t]
\centering
{
\small
\setlength{\tabcolsep}{2.0pt}
\begin{tabular}{@{}lrrrrrrrrrr@{}}
\toprule
& \multicolumn{5}{c}{Amazon Beauty}
& \multicolumn{5}{c}{Yelp} \\
\cmidrule(lr){2-6}\cmidrule(l){7-11}
Method & H@1 & H@5 & H@10 & N@5 & N@10
       & H@1 & H@5 & H@10 & N@5 & N@10 \\
\midrule
LightGCN      & 0.00742 & 0.02759 & 0.04396 & 0.01738 & 0.02262
              & \underline{0.00429} & 0.01738 & 0.03091 & 0.01075 & 0.01507 \\
SASRec        & 0.00782 & 0.02942 & 0.04914 & 0.01869 & 0.02503
              & 0.00380 & 0.01686 & 0.02979 & 0.01028 & 0.01444 \\
LETTER-LC-Rec & 0.01082 & 0.02705 & 0.03926 & 0.01903 & 0.02297
              & 0.00284 & 0.00998 & 0.01724 & 0.00638 & 0.00871 \\
LETTER-TIGER  & \underline{0.01145} & \underline{0.03716} & \underline{0.05755} & \underline{0.02441} & \underline{0.03100}
              & \textbf{0.00540} & \textbf{0.02043} & \textbf{0.03392} & \textbf{0.01283} & \textbf{0.01714} \\
GraphRfi      & 0.00284 & 0.01335 & 0.02557 & 0.00805 & 0.01193
              & 0.00285 & 0.01113 & 0.01964 & 0.00700 & 0.00972 \\
PGT4Rec       & 0.00410 & 0.01858 & 0.03030 & 0.01121 & 0.01497
              & 0.00275 & 0.01107 & 0.01935 & 0.00694 & 0.00958 \\
ReliGRec      & \textbf{0.01771} & \textbf{0.04431} & \textbf{0.06269} & \textbf{0.03107} & \textbf{0.03698}
              & \textbf{0.00540} & \underline{0.02010} & \underline{0.03388} & \underline{0.01259} & \underline{0.01701} \\
\bottomrule
\end{tabular}
}
\captionsetup{labelsep=colon}
\caption{Overall next-item recommendation performance. The best result is
shown in bold, and the second-best result is underlined.}
\label{tab:overall_results}
\end{table*}

\subsection{Weak-Risk Estimation (RQ2)}

Table~\ref{tab:risk_results}(a) compares ReliGRec with GraphRfi and PGT4Rec on
the user-disjoint weak-risk test sets. ReliGRec achieves the best AUPRC, AUROC,
and F1 on both datasets. Compared with the strongest baseline for each ranking
metric, its AUPRC improves from 0.2187 to 0.2633 on Beauty and from 0.1990 to
0.2990 on Yelp, while its AUROC improves from 0.4897 to 0.5818 and from 0.5311
to 0.6550, respectively. Panel (b) compares single-view and fused weak-risk
inputs on Beauty. Behavior-only improves AUPRC and AUROC over Graph-only
(0.2484 and 0.5586 versus 0.2090 and 0.5126), whereas Graph-only yields a
higher F1 at the fixed cutoff (0.3471 versus 0.3318). Combining both views
achieves the best AUPRC, AUROC, and F1 (0.2631, 0.5817, and 0.3651),
supporting their complementarity. All metrics quantify agreement with
user-level weak-risk proxy labels rather than verified fraud-detection
performance.

\begin{table}[t]
\centering
{
\small
\setlength{\tabcolsep}{2.7pt}
\begin{tabular}{@{}llrrr@{}}
\toprule
\multicolumn{5}{l}{\textit{(a) Overall weak-risk estimation}}\\
Dataset & Method & AUPRC & AUROC & F1\\
\midrule
Beauty & GraphRfi & \underline{0.2187} & 0.4778 & \underline{0.3096}\\
       & PGT4Rec  & 0.2068 & \underline{0.4897} & 0.2908\\
       & ReliGRec & \textbf{0.2633} & \textbf{0.5818} & \textbf{0.3653}\\
\cmidrule(lr){1-5}
Yelp   & GraphRfi & 0.1975 & 0.5124 & \underline{0.3202}\\
       & PGT4Rec  & \underline{0.1990} & \underline{0.5311} & 0.2010\\
       & ReliGRec & \textbf{0.2990} & \textbf{0.6550} & \textbf{0.3645}\\
\midrule
\multicolumn{5}{l}{\textit{(b) Weak-risk-input analysis on Beauty}}\\
\multicolumn{2}{l}{Weak-risk input} & AUPRC & AUROC & F1\\
\cmidrule(lr){1-5}
\multicolumn{2}{l}{Behavior only}
  & \underline{0.2484} & \underline{0.5586} & 0.3318\\
\multicolumn{2}{l}{Graph only}
  & 0.2090 & 0.5126 & \underline{0.3471}\\
\multicolumn{2}{l}{Graph + Behavior}
  & \textbf{0.2631} & \textbf{0.5817} & \textbf{0.3651}\\
\bottomrule
\end{tabular}
}
\caption{User-level weak-risk proxy prediction. (a) Overall comparison on
protocol-matched weak-risk test users. (b) Weak-risk-input analysis on Beauty. AUPRC is
the primary metric; F1 uses the fixed cutoff \(\tau=0.5\). The best result is
shown in bold and the second-best is underlined.}
\label{tab:risk_results}
\end{table}

\subsection{Prompt-Routing Analysis (RQ3)}

We compare four routing policies on Beauty using the same checkpoint, users,
candidate catalog, and decoding configuration. All-Simple and All-Cautious
apply one prompt to every request. Random matched-ratio assigns exactly the
same number of requests to the Cautious Prompt as Weak-Risk-Guided, and is
repeated 500 times using cached per-user predictions. Proxy-deviation groups
contain only held-out labeled users.

\begin{table}[t]
\centering
{
\small
\setlength{\tabcolsep}{1.8pt}
\begin{tabular}{@{}lccccc@{}}
\toprule
Policy
& H@10
& N@10
& \shortstack{Lower\\N@10}
& \shortstack{Higher\\N@10}
& \shortstack{Cautious\\(\%)} \\
\midrule
All-Simple
& \textbf{0.0642} & \textbf{0.0379} & \textbf{0.0337}
& \textbf{0.0462} & 0.0 \\
All-Cautious
& 0.0616 & 0.0360 & 0.0322 & 0.0384 & 100.0 \\
Random
& 0.0625 & 0.0366 & \underline{0.0327} & 0.0411 & 66.9 \\
Weak-Risk
& \underline{0.0627} & \underline{0.0370} & 0.0321
& \underline{0.0417} & 66.9 \\
\bottomrule
\end{tabular}
}
\caption{Prompt-routing results on Beauty. Weak-Risk denotes
weak-risk-guided routing. Lower and Higher refer to the held-out lower- and
higher-proxy-deviation groups. Bold and underlined values indicate the best
and second-best results, respectively.}
\label{tab:routing_results}
\end{table}

For Random routing, the 95\% confidence intervals are
[0.0616, 0.0633] for H@10, [0.0362, 0.0370] for N@10,
[0.0309, 0.0343] for lower-proxy-deviation N@10, and
[0.0377, 0.0448] for higher-proxy-deviation N@10.

The weak-risk-guided policy is numerically higher than All-Cautious in
NDCG@10 (0.0370 versus
0.0360), but All-Simple remains stronger at 0.0379. Its overall NDCG@10 and the
NDCG@10 for the higher-proxy-deviation group also fall within the 95\%
intervals of matched-ratio random routing. The measured end-to-end wall-clock
costs are 0.460, 0.548, and 0.479 seconds per user for All-Simple,
All-Cautious, and the weak-risk-guided policy, respectively. Thus, the current
evidence establishes weak-risk-guided prompt selection, but does not show that
the estimated risk yields a recommendation improvement over the strongest
uniform policy or non-informative matched routing. Because the score is learned
with positive-class weighting and is not calibrated, the 66.9\% Cautious-route
rate should not be interpreted as the prevalence of higher proxy deviation.

\subsection{Graph-Conditioning and Temporal-Aggregation Ablation (RQ4)}

To isolate the contribution of temporal Graph Tokens to generation, we compare
the graph-conditioned model with a history-only generator while keeping the
Semantic IDs, language-model backbone, training budget, beam size, prompt
policy, and evaluation procedure unchanged. As shown in
Table~\ref{tab:graph_conditioning}, temporal graph conditioning improves all
four metrics on Beauty. Relative to the history-only generator, Hit@5 and
Hit@10 increase by 2.78\% and 1.58\%, while NDCG@5 and NDCG@10 increase by
4.65\% and 3.84\%, respectively. We further replace the temporal sequence with
a one-snapshot static Graph Token and retrain the model under the same
protocol. Under this seed, the static variant also exceeds History-only,
attaining the best Hit@5, NDCG@5, and NDCG@10, while the temporal variant
attains the best Hit@10. Thus, both graph-conditioned variants improve on the
history-only generator in this controlled run, but temporal aggregation does
not uniformly dominate the static representation on ranking metrics. For
user-level weak-risk estimation, the temporal variant yields higher AUPRC
(0.2631 versus
0.2313), AUROC (0.5817 versus 0.5299), and F1 (0.3651 versus 0.2826),
suggesting that time-resolved collaborative evidence is more informative for
proxy-label prediction in this run.

\begin{table}[t]
\centering
{
\small
\setlength{\tabcolsep}{4.2pt}
\begin{tabular}{lrrrr}
\toprule
Generator & H@5 & H@10 & N@5 & N@10 \\
\midrule
History only          & 0.0432 & 0.0632 & 0.0301 & 0.0365 \\
+ Static Graph Token  & \textbf{0.0449} & 0.0634 & \textbf{0.0324} & \textbf{0.0384} \\
+ Temporal Graph Token & 0.0444 & \textbf{0.0642} & 0.0315 & 0.0379 \\
\bottomrule
\end{tabular}
}
\caption{Effect of graph conditioning and temporal aggregation on Beauty
(seed 42).}
\label{tab:graph_conditioning}
\end{table}

\section{Related Work}
\subsection{LLM-based Generative Recommendation}

Semantic-ID-based generative recommendation represents items as discrete codes.
TIGER predicts tuple-structured Semantic IDs, while LC-Rec and LETTER enhance
item tokenization through semantic--collaborative alignment and hierarchical
regularization~\citep{rajput2023tiger,zheng2024lcrec,wang2024letter}. CoLLM,
CLLM4Rec, GraphLoRA, and TCA4Rec further incorporate collaborative signals
through input projection, vocabulary expansion, parameter adaptation, or
token-level alignment~\citep{zhang2025collm,zhu2024cllm4rec,mu2026graphlora,
lin2026tca4rec}. SASRec and TGSRec provide foundations for sequential and
temporal collaborative modeling~\citep{kang2018sasrec,fan2021tgsrec}. These
methods mainly improve item representation and collaborative conditioning, but
rarely adapt the generation policy to user-level risk before decoding.

\subsection{Robust and Risk-Aware Recommendation}

Robust recommenders incorporate risk signals through joint recommendation and
detection, graph aggregation, user reweighting, or representation rectification,
as exemplified by GraphRfi, PGT4Rec, Trust-GRS, LoRec, and DITaR
~\citep{zhang2020graphrfi,nguyen2025pgt4rec,mu2025trustgrs,zhang2024lorec,
qin2026ditar}. These methods mainly intervene during training. By contrast,
GUIDER and UGR exploit generation uncertainty for reranking, diversity control,
rejection, or truncation~\citep{xu2026guider,fan2026ugr}, operating during or
after generation. ReliGRec instead estimates user-level weak risk from
sequential behavior and temporal collaborative context before decoding, using
it to route users between Simple and Cautious Prompts while preserving Graph
Token conditioning.

\section{Conclusion}
In this paper, we propose ReliGRec, a risk-aware generative recommendation
framework that integrates collaborative modeling, user-level weak-risk
estimation, and weak-risk-guided prompt routing. To capture evolving
collaborative relations, ReliGRec constructs temporal Graph Tokens from
interaction snapshots and uses their aggregated representation to condition
next-item Semantic ID generation. The temporal Graph Tokens and a Behavior
Token encoding sequential preferences provide complementary collaborative and
behavioral evidence for user-level weak-risk estimation with partially labeled
users.
Based on the predicted weak-risk score, ReliGRec selects between the Simple and
Cautious Prompts before generation. The reported results show competitive
recommendation and proxy-label prediction, while the controlled Beauty
analysis supports the contribution of graph conditioning. The routing results
characterize group-dependent changes in recommendation quality and inference
cost, but do not establish an advantage over All-Simple or matched-ratio random
routing.

\label{tmp:body-end}
\bibliography{references}

\end{document}